\documentclass[12pt,a4paper,oneside,onecolumn]{article}
\usepackage[cp1251]{inputenc}
\usepackage[english]{babel}
\usepackage{amsmath,amssymb}
\usepackage{graphicx}
\usepackage[square,comma,numbers,sort&compress]{natbib}

\begin{document}
\begin{flushleft}
Bulletin of the Lebedev Physics Institute, 2013, Vol. 40, No. 4, pp. 91--96.

\copyright\ Allerton Press, Inc., 2013.
\footnote{Original Russian Text \copyright\ D. N. Sob'yanin, 2013, published in Kratkie Soobshcheniya po Fizike, 2013, Vol. 40,\\No. 4, pp. 15--24.}
\end{flushleft}
\vspace{0.5\baselineskip}
\begin{center}
\textbf{\large Theory of Bose-Einstein Condensation of Light in a Microcavity}

D. N. Sob'yanin

Lebedev Physical Institute, Russian Academy of Sciences,\\
Leninskii pr. 53, Moscow, 119991 Russia; e-mail: sobyanin@lpi.ru

Received December 5, 2012

\vspace{\baselineskip}
\textbf{Abstract}
\end{center}

A theory of Bose-Einstein condensation (BEC) of light in a dye microcavity is developed. The photon polarization degeneracy and the interaction between dye molecules and photons in all of the cavity modes are taken into account. The theory goes beyond the grand canonical approximation and allows one to determine the statistical properties of the photon gas for all numbers of dye molecules and photons at all temperatures, thus describing the microscopic, mesoscopic, and macroscopic light BEC from a general perspective. A universal relation between the degrees of second-order coherence for the photon condensate and the polarized photon condensate is obtained. The photon Bose-Einstein condensate can be used as a new source of nonclassical light.

\vspace{\baselineskip}\noindent
\textbf{Keywords:} Bose-Einstein condensation of light, condensate fluctuations, sub-Poissonian statistics, nonclassical light

\newpage
Bose-Einstein condensation (BEC) of light in a dye microcavity has been observed recently~\cite{KlaersEtal2010}. In this experiment, photons are confined in a bispherical optical microcavity filled with a dye solution. In the microcavity, the continuous absorption and reemission of photons by dye molecules occurs. Since the free spectral range of the microcavity is comparable to the spectral width of the dye, the emission of photons with a fixed longitudinal quantum number dominates. In addition, the curvature of the mirrors induces a harmonic trapping potential for photons. Thus, we deal with a two-dimensional massive photon gas in a harmonic potential, and the photon gas continuously interacts with the dye solution.

The aim of this paper is to present a consistent theory of the light BEC. This theory takes account of the photon polarization degeneracy and the interaction of all the cavity modes with dye molecules. The grand canonical approximation is never used. The theory allows one to fully determine the statistical properties of the photon gas without restrictions on the numbers of photons and dye molecules and, in particular, to study fluctuations of the whole photon gas, photon condensate, and polarized photon condensate. Moreover, it predicts unusual nonclassical intensity correlations that have not been observed previously in the light BEC experiment.

In the optical microcavity, the longitudinal quantum number is fixed. The microcavity has a discrete equidistant energy spectrum such that the energy and degeneracy of the $m$th level are $\hbar\omega_m=\hbar\omega_0+m\hbar\Omega$ and $g_m=2(m+1)$, respectively, where $\hbar\omega_0$ is the energy of a ground-mode photon and $\Omega$ is the trapping frequency. If the longitudinal quantum number $q_0$, mirror spacing $D_0$, mirror radius of curvature $R$, and linear refraction index of the dye solution $n_0$ are known, then $\hbar\omega_0=m_\text{ph}c_0^2+\hbar\Omega$ and $\Omega=c_0\sqrt{2/D_0R}$, where $m_\text{ph}=\pi\hbar q_0/c_0 D_0$ is the photon mass and $c_0=c/n_0$ is the speed of light in the medium. It is convenient to number all the modes with the same energy. Then we can characterize every mode by two quantum numbers: the number~$m$ of the energy level and the number~$i$ of the mode. Given~$m$, $0\leqslant i\leqslant g_m-1$.

A ground-mode photon is defined as a photon with energy~$\hbar\omega_0$, and the photon condensate is a set of ground-mode photons. Because of the twofold polarization degeneracy of the ground state, we additionally define a polarized ground-mode photon, a ground-mode photon of a definite polarization, and consider a polarized photon condensate, a set of polarized ground-mode photons. Then the whole photon condensate is a mixture of two polarized photon condensates, the first comprising mode-$00$ photons and the second comprising mode-$01$ photons. Each of the two modes corresponds to one of the two photon polarizations, and the two polarized condensates have the same statistical properties.

Let $n_{mi}$ be the number of mode-$mi$ photons and $|\text{ph}\rangle$ be a Fock state of the photon gas, which is given by a vector composed of the numbers of photons in each mode: $|\text{ph}\rangle=|\{n_{mi}\}\rangle$. Let us denote the total number of photons for this state by~$n_{|\text{ph}\rangle}$. The entropy of the photon gas in a Fock state is zero, $s^\text{ph}_{|\text{ph}\rangle}=0$, and the total energy is $u^\text{ph}_{|\text{ph}\rangle}=n_{|\text{ph}\rangle}\hbar\omega_0+u^\perp_{|\text{ph}\rangle}$, where $u^\perp_{|\text{ph}\rangle}=\hbar\Omega\varepsilon_{|\text{ph}\rangle}$ is the transverse energy and $\varepsilon_{|\text{ph}\rangle}=\sum n_{mi}m$ is the reduced transverse energy.

The photon gas exchanges excitations with dye molecules so that the total number $n_\Sigma$ of excitations (photons and excited dye molecules) is fixed. The statistical properties of the photon gas are determined by a probability distribution $\{\pi_{|\text{ph}\rangle}\}$, where $\pi_{|\text{ph}\rangle}$ is the probability of state $|\text{ph}\rangle$.

To find this distribution, I will use the hierarchical maximum entropy principle, which has been developed very recently in the context of generalized superstatistics~\cite{Sobyanin2012}.  This statistics describes generalized superstatistical systems, the essential feature of which is the sufficient time-scale separation between different levels of the system dynamics~\cite{Sobyanin2011}. The hierarchical maximum entropy principle consists in arranging the levels of dynamics in increasing order of dynamical time scale and consecutively maximizing the entropy at each level.

The whole system is composed of two subsystems: the subsystem of the dye solution and the subsystem of the photon gas. The subsystem of the dye solution in turn consists of a solvent and $n_\text{d}$ dye molecules, among which there are $n_\text{d}-n_\Sigma+n_{|\text{ph}\rangle}$ ground-state molecules and $n_\Sigma-n_{|\text{ph}\rangle}$ excited molecules. Each molecule is in contact with the solvent, which plays the role of thermostat. The population of the electronic states of dye molecules is quickly thermalized, with the characteristic time $\sim1$~ps at room temperature \cite{Shank1975,Schaefer1990,Lakowicz2006}. Since the typical fluorescence lifetime is $\sim1{-}10$~ns, the emission of photons occurs from thermally equilibrated excited states. This apparent time-scale separation allows us to directly apply the hierarchical maximum entropy principle to this system.

Consider the subsystem of all the dye molecules. Let $g_0(\varepsilon_0)$ and $g_1(\varepsilon_1)$ be the density of rovibrational states for the ground, $S_0$, and first excited, $S_1$, singlet electronic state, respectively. Note that $\varepsilon_i=E-E_i$, where $E_i$ is the lowest-energy substate of $S_i$, $i=0,1$, and $g_i(\varepsilon_i)=0$ for any $\varepsilon_i<0$. Maximizing the entropy at the lower dynamical level, corresponding to fast thermalization, allows us to obtain the entropy $s^\text{d}_{|\text{ph}\rangle}=(n_\text{d}-n_\Sigma+n_{|\text{ph}\rangle})s_0+(n_\Sigma-n_{|\text{ph}\rangle})s_1+\ln\bigl(
\begin{smallmatrix}
n_\text{d}\\n_\Sigma-n_{|\text{ph}\rangle}
\end{smallmatrix}
\bigr)$ and mean energy $u^\text{d}_{|\text{ph}\rangle}=(n_\text{d}-n_\Sigma+n_{|\text{ph}\rangle})u_0+(n_\Sigma-n_{|\text{ph}\rangle})u_1$ of the subsystem~\cite{Sobyanin2012}. Here, $s_i=\ln z_i+\beta(u_i-E_i)$ and $u_i=E_i-z^{-1}_i d z_i/d\beta$ are the entropy and mean energy of a dye molecule, with $i=0$ for a ground-state molecule and $i=1$ for an excited molecule, $z_i=\int e^{-\beta\varepsilon}g_i(\varepsilon)d\varepsilon$ is the reduced partition function, and $\beta=(k_B T)^{-1}$ is the inverse temperature.

Now consider the system as a whole. For a fixed $|\text{ph}\rangle$, the entropy and mean energy of the system are given by $S_{|\text{ph}\rangle}=s^\text{d}_{|\text{ph}\rangle}+s^\text{ph}_{|\text{ph}\rangle}$ and $U_{|\text{ph}\rangle}=u^\text{d}_{|\text{ph}\rangle}+u^\text{ph}_{|\text{ph}\rangle}$, respectively. Maximizing the entropy (cf. Ref.~\cite{Sobyanin2012})
$S=-\sum\pi_{|\text{ph}\rangle}\ln\pi_{|\text{ph}\rangle}
+\sum\pi_{|\text{ph}\rangle}S_{|\text{ph}\rangle}$
under the normalization condition $\sum\pi_{|\text{ph}\rangle}=1$ and the mean energy constraint $\sum\pi_{|\text{ph}\rangle}U_{|\text{ph}\rangle}=U$, with $\beta$ as the Lagrange multiplier corresponding to the latter constraint, yields the probability of state $|\text{ph}\rangle$ in the form
\begin{eqnarray}
\frac{\pi_{|\text{ph}\rangle}}{\pi_0}&=&
\begin{pmatrix}
n_\text{d}\\n_\Sigma
\end{pmatrix}
^{-1}
\begin{pmatrix}
n_\text{d}\\n_\Sigma-n_{|\text{ph}\rangle}
\end{pmatrix}
\biggl(\frac{z_0}{z_1}\biggr)^{n_{|\text{ph}\rangle}}\nonumber\\
& &\times\exp\{-\beta[n_{|\text{ph}\rangle}\hbar(\omega_0-\omega_\text{e})+u^\perp_{|\text{ph}\rangle}]\},\label{piPhOverPi0}
\end{eqnarray}
where $\pi_0\equiv\pi_{|0\rangle}$ is the probability that there are no photons in the optical microcavity and $\hbar\omega_\text{e}=E_1-E_0$.

It is possible to write the master equation that describes the interaction between photons and dye molecules and to directly check that the probability distribution $\{\pi_{|\text{ph}\rangle}\}$ determined with the hierarchical maximum entropy principle is the stationary solution of the master equation. This fact will be shown in a separate paper. In a simplified case, this fact was shown in Ref.~\cite{Sobyanin2012}.

Equation \eqref{piPhOverPi0} determines all statistical characteristics of the photon gas. In what follows, we will confine our attention to fluctuations of the photon condensate, polarized photon condensate, and photon gas as a whole.

First consider fluctuations of the whole photon gas. Denote by $|\text{ph}^n\rangle$ a state with $n$ photons: $n_{|\text{ph}^n\rangle}=n$. The probability that the photon gas comprises $n$ photons is $\pi_n=\sum\pi_{|\text{ph}^n\rangle}$, and hence
\begin{equation}
\label{piNOverPi0}
\frac{\pi_n}{\pi_0}=
\begin{pmatrix}
n_\text{d}\\n_\Sigma
\end{pmatrix}
^{-1}
\begin{pmatrix}
n_\text{d}\\n_\Sigma-n
\end{pmatrix}
r^n a_n,
\end{equation}
where $r=(z_0/z_1)\exp[-\beta\hbar(\omega_0-\omega_\text{e})]$, $a_n=\sum q^{\varepsilon_{|\text{ph}^n\rangle}}$, and $q=\exp(-\beta\hbar\Omega)$. The quantity $a_n$ is determined from the recursive relations: $a_0=1$ and $a_n=n^{-1}\sum_{m=1}^n c_m a_{n-m}$, $n\geqslant1$, where $c_n=2(1-q^n)^{-2}$. Equation~\eqref{piNOverPi0} allows us to find $\pi_0=(\sum_{n=0}^{n_\Sigma}\pi_n/\pi_0)^{-1}$ and then calculate $\pi_n$ for all positive $n\leqslant n_\Sigma$.

Next consider fluctuations of the polarized photon condensate. The probability $\pi^{00}_m$ that the polarized condensate comprises $m$ photons is obtained by summing $\pi_{|\text{ph}\rangle}$ over all states with $m$ mode-$00$ photons:
\begin{equation}
\label{pi00m}
\frac{\pi^{00}_m}{\pi_0}=
\begin{pmatrix}
n_\text{d}\\n_\Sigma
\end{pmatrix}
^{-1}
\sum_{n=m}^{n_\Sigma}
\begin{pmatrix}
n_\text{d}\\n_\Sigma-n
\end{pmatrix}
r^n a^{00}_{n-m},
\end{equation}
where $a^{00}_0=1$ and $a^{00}_n=a_n-a_{n-1}$, $n\geqslant1$.

Finally consider fluctuations of the whole photon condensate. The probability $\pi^0_m$ that the condensate comprises $m$ photons is obtained by summing $\pi_{|\text{ph}\rangle}$ over all states with $m$ ground-mode photons:
\begin{equation}
\label{pi0m}
\frac{\pi^0_m}{\pi_0}=
\begin{pmatrix}
n_\text{d}\\n_\Sigma
\end{pmatrix}
^{-1}
(m+1)
\sum_{n=m}^{n_\Sigma}
\begin{pmatrix}
n_\text{d}\\n_\Sigma-n
\end{pmatrix}
r^n a^0_{n-m},
\end{equation}
where $a^0_0=1$ and $a^0_n=a^{00}_n-a^{00}_{n-1}$, $n\geqslant1$.

Now consider the situation where the mean photon number is fixed whereas the temperature is varied. As an example, choose the ratio of the reduced partition functions $z_1/z_0=1$, the reduced detuning $(\omega_0-\omega_e)/\Omega=-10^3$, and the number of dye molecules $n_d=10^6$, and study how the main characteristics of the photon gas change with temperature for the series of mean photon numbers $\langle n\rangle=1$, $2$, $4$, $10$, $10^2$, $10^3$, and $10^4$.

\begin{figure}
\includegraphics[width=17cm]{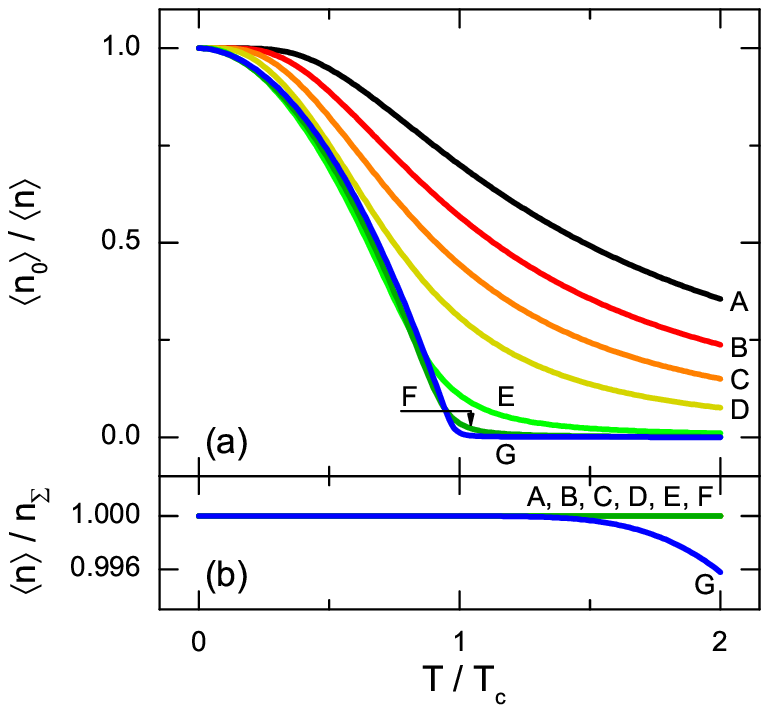}
\caption{\label{Fig1}(a) Condensate fraction $\langle n_0\rangle/\langle n\rangle$ and (b) photon fraction $\langle n\rangle/n_\Sigma$ against reduced temperature $T/T_c$. Parameters used: ratio of reduced partition functions $z_1/z_0=1$; reduced detuning $(\omega_0-\omega_e)/\Omega=-10^3$; dye molecule number $n_d=10^6$; mean photon number $\langle n\rangle=1$ (curve~A), $2$ (B), $4$ (C), $10$ (D), $10^2$ (E), $10^3$ (F), and $10^4$ (G).}
\end{figure}

Figure~\ref{Fig1}(a) shows the dependence of the condensate fraction $\langle n_0\rangle/\langle n\rangle$ on the reduced temperature $T/T_c$, where $\langle n_0\rangle$ is the mean number of ground-mode photons and $T_c = \hbar\Omega\sqrt{3\langle n\rangle}/\pi k_B$ is the critical temperature. As the mean photon number increases, the condensate fraction approaches the parabolic dependence
$\langle n_0\rangle/\langle n\rangle=1-(T/T_c)^2$, $T\leqslant T_c$, and $\langle n_0\rangle/\langle n\rangle=0$, $T>T_c$.
Thus, as temperature decreases, the photon gas undergoes BEC, and the condensate photon number becomes a macroscopic fraction of the total photon number.

Figure~\ref{Fig1}(b) shows the temperature dependence of the photon fraction $\langle n\rangle/n_\Sigma$. We see that, for the parameters considered, excitations in the optical cavity are mainly photon excitations.

Define the quantum degree of second-order coherence (the normalized zero-delay second-order correlation function) \cite{Glauber1963,Glauber1963b,Loudon2000} for the photon condensate, $g^{(2)}_0(0)=\langle n_0(n_0-1)\rangle/\langle n_0\rangle^2$, and for the polarized photon condensate, $g^{(2)}_{00}(0)=\langle n_{00}(n_{00}-1)\rangle/\langle n_{00}\rangle^2$, where $n_0$ and $n_{00}$ are the fluctuating photon numbers corresponding to these condensates. A universal relation
\begin{equation}
\label{universalG20Relation}
g^{(2)}_0(0)=\frac{3}{4}\,g^{(2)}_{00}(0)
\end{equation}
follows from Eqs.~\eqref{pi00m} and~\eqref{pi0m}. This relation holds for all numbers of dye molecules and photons at all temperatures.

\begin{figure}
\includegraphics[width=17cm]{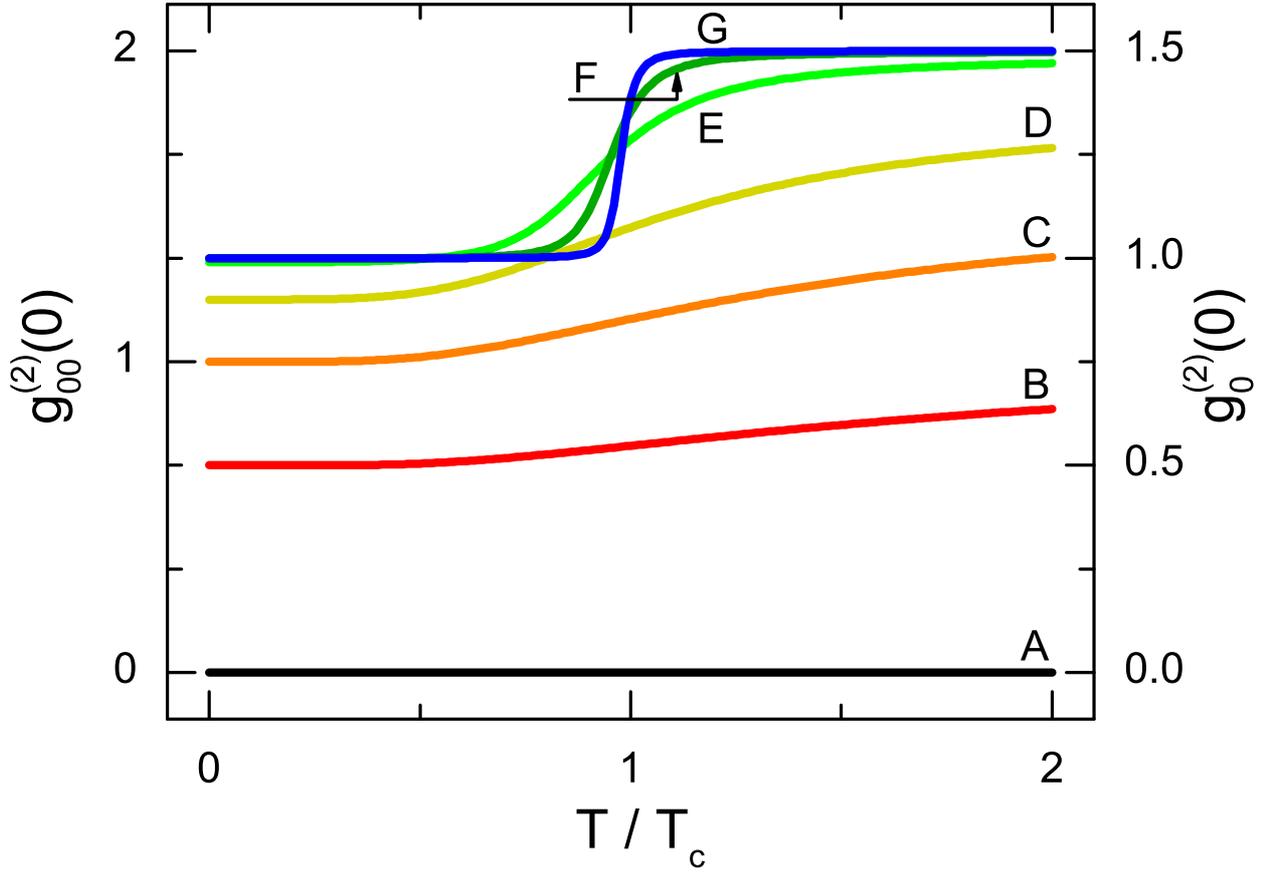}
\caption{\label{Fig2}Degree of second-order coherence for (left axis) polarized, $g^{(2)}_{00}(0)$, and (right axis) whole, $g^{(2)}_{0}(0)$, photon condensate against reduced temperature $T/T_c$. Parameters used are the same as in Fig.~\ref{Fig1}. Universal relation~\eqref{universalG20Relation} is seen.}
\end{figure}

Figure~\ref{Fig2} shows the temperature dependence of $g^{(2)}_{00}(0)$ and~$g^{(2)}_0(0)$. Each curve corresponds both to $g^{(2)}_{00}(0)$ (left axis) and to $g^{(2)}_0(0)$ (right axis), illustrating the universal relation~\eqref{universalG20Relation}. For the polarized photon condensate, we observe Bose-Einstein statistics at high temperatures, $g^{(2)}_{00}(0)=2$, and super-Poissonian statistics even at zero temperature, $g^{(2)}_{00}(0)=4/3$, when the mean photon number is large. For the whole photon condensate, we observe at first glance Poissonian but in fact sub-Poissonian (see below) statistics at low temperatures, $g^{(2)}_0(0)\approx1$ but $<1$, and super-Poissonian statistics at high temperatures, $g^{(2)}_0(0)=3/2$.

\begin{figure}
\includegraphics[width=17cm]{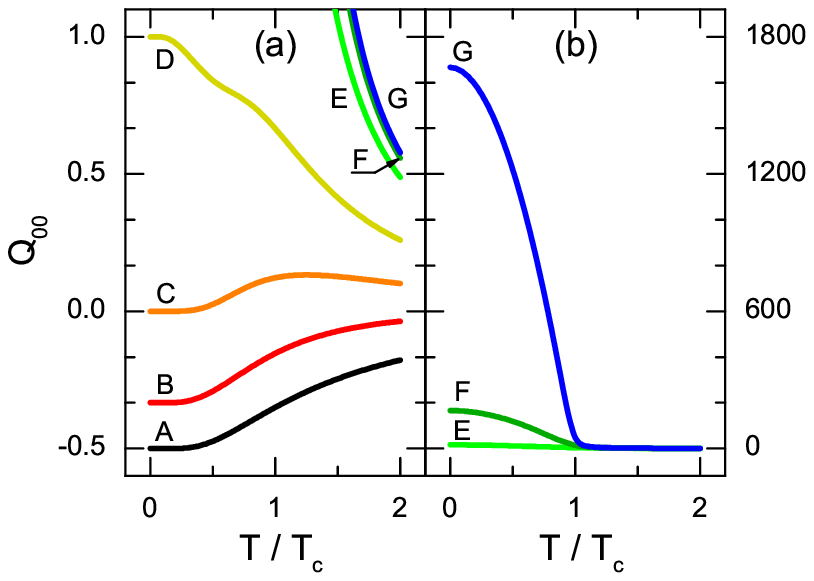}
\caption{\label{Fig3}Mandel parameter $Q_{00}$ for polarized photon condensate against reduced temperature $T/T_c$. Parameters used are the same as in Fig.~\ref{Fig1}.}
\end{figure}

\begin{figure}
\includegraphics[width=17cm]{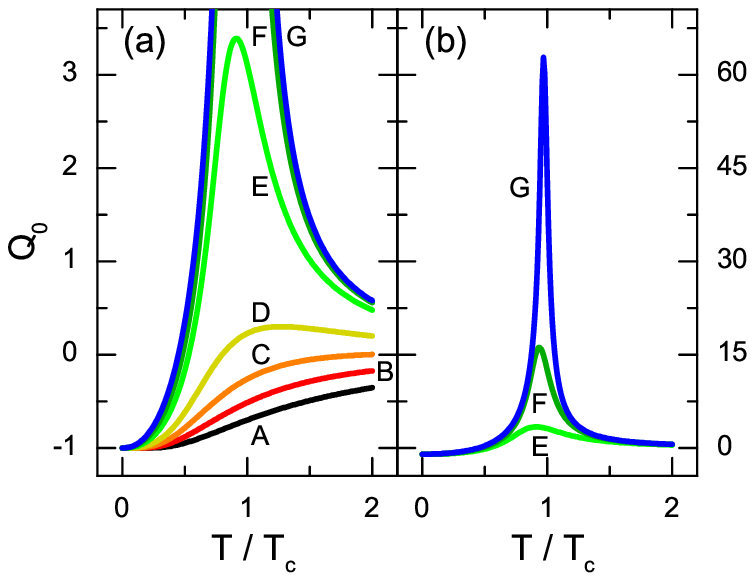}
\caption{\label{Fig4}Mandel parameter $Q_0$ for whole photon condensate against reduced temperature $T/T_c$. Parameters used are the same as in Fig.~\ref{Fig1}.}
\end{figure}

Significantly, when the mean photon number is small, we observe sub-Poissonian statistics, both for the polarized, $g^{(2)}_{00}(0)<1$,, and for the whole, $g^{(2)}_0(0)<1$, photon condensate, and even at temperatures higher than the critical temperature (see curves A and B in Fig.~\ref{Fig2}). The same is demonstrated in Figs.~\ref{Fig3} and~\ref{Fig4}, which show, respectively, the Mandel parameter \cite{Mandel1979} for the polarized, $Q_{00}=(g^{(2)}_{00}(0)-1)\langle n_{00}\rangle$, and for the whole, $Q_0=(g^{(2)}_0(0)-1)\langle n_0\rangle$, photon condensate: both $Q_{00}<0$ and $Q_0<0$ for curves~A and~B in Figs.~\ref{Fig3}(a) and~\ref{Fig4}(a). As the mean photon number increases, there appear the states for which the statistics of the photon condensate is sub-Poissonian ($g^{(2)}_0(0)<1$, $Q_0<0$) at some temperature, whereas the statistics of the polarized photon condensate is Poissonian ($g^{(2)}_{00}(0)=1$, $Q_{00}=0$) or super-Poissonian ($g^{(2)}_{00}(0)>1$, $Q_{00}>0$); curves~C and~D in Figs.~\ref{Fig2}, \ref{Fig3}(a), and~\ref{Fig4}(a) at low temperatures illustrate these two cases.

Figure~4(a) shows that the Mandel parameter $Q_0$ for the photon condensate tends to $-1$ as temperature vanishes. This temperature dependence of~$Q_0$ again reflects the light BEC: the photon condensate as a whole is in a Fock state at zero temperature for any mean photon number. The condensate photon number does not fluctuate and equals the total photon number, which coincides with the total excitation number. Therefore, the photon statistics for the whole condensate is always sub-Poissonian at zero temperature, with $g^{(2)}_0(0)=1-n_0^{-1}<1$. The polarized photon condensate, however, fluctuates at zero temperature, with $Q_{00}\geqslant-1/2>-1$ in Fig.~\ref{Fig3}(a), because $n_0$ ground-mode photons are distributed between the two polarization states in $n_0+1$ ways with equal probability.

For large photon numbers, $Q_{00}$ is relatively small above the critical temperature, but sharply increases as temperature passes down through the critical value, thereby indicating the beginning of BEC [see Fig.~\ref{Fig3}(b)]. In turn, $Q_0$ is small both at high and at low temperatures, but has a sharp peak near the critical temperature, in the region where the photon statistics changes [see Fig.~\ref{Fig4}(b)].

In conclusion, I have developed the theory of Bose-Einstein condensation of light in a dye microcavity. I have taken into account the photon polarization degeneracy and the interaction of photons in all the cavity modes with dye molecules. The theory describes the microscopic, mesoscopic, and macroscopic light BEC from a general perspective and allows one to find all statistical characteristics of the photon gas for all numbers of dye molecules and photons at all temperatures. In particular, it allows one to study fluctuations of the whole photon gas, photon condensate, and polarized photon condensate. I have obtained the universal relation between the degrees of second-order coherence for the polarized photon condensate and the whole photon condensate. The theory predicts sub-Poissonian photon statistics for these condensates in the case of small mean photon numbers even for temperatures above the critical temperature. At low temperatures, the photon statistics of the whole condensate is sub-Poissonian even for large mean photon numbers. The behavior of the degree of second-order coherence for the polarized and the whole photon condensates implies that a dye microcavity in which the light BEC takes place can be the source of nonclassical light, both polarized and unpolarized, with sub-Poissonian photon statistics and antibunching.

I propose to observe experimentally the predictions of the theory, viz., sub-Poissonian statistics and the universal relation~\eqref{universalG20Relation}.

\clearpage
\clearpage
\clearpage
\clearpage
\end{document}